# MICROFOUNDATIONS OF EXPECTED UTILITY AND RESPONSE-TIME[1]

Gonzalo Valdes-Edwards[1] and Salvador Valdes-Prieto[2]

[1] Andres Bello University Institute for Public Policy, Fernandez Concha 700, Las Condes, Region Metropolitana, 759 1538, Chile. gonzalo.valdes.e@unab.cl, +56 9 8501 2204.

[2] Adolfo Ibanez University School of Business, Diagonal Las Torres 2640, Room 227A, Peñalolen, Metropolitan Region, 794 1169, Chile. salvador.valdes.p@uai.cl , +56 2 2933 0408.

Dato of this version: February 18, 2023

**Abstract**

This paper builds a rule for decision-making from the physical behavior of single neurons, the well-established neural circuitry of mutual inhibition, and the evolutionary principle of natural selection. No axioms are used in the derivation of this rule.

The paper provides a microfoundation to both Economics Choice Theory and Cognitive Psychology's Response-Times Theory.

The paper finds how classical expected utility should be modified to account for much neuroscientific evidence, and how neuroscientific correlates of choice should be linked to utility. In particular, the model implies the concept of utility is a network property and cannot be calculated as a function of frequencies in one layer of neurons alone; it requires understanding how different layers work together.

The resulting rule is simple enough to model markets and games as is customary in the social sciences. Utility maximization and inaction are endogenous to the model, cardinality and independence of irrelevant alternatives -properties present in classical and random utility theories- are only met in limiting situations.

---

* The first stage of this paper benefited from comments from Mark Machina, Andrei Shleifer, Sam Gershman, Nicola Gennaioli, Marciano Siniscalchi, Philip L. Smith, Carlos Rodríguez-Sickert, Ricardo Guzmán, Ricardo Paredes, Alain Raymond, Julio C. Saavedra, René San Martín, Paul Wander, and from participants at the Theory Seminars at the Catholic University of Chile and Adolfo Ibañez University in 2015. Initial versions appeared as Working Papers, with a different title. Substantial further work was added during the Covid-19 pandemic. This research has not received any specific grant from agencies in the public, commercial, or non-profit sectors.

## *1. Introduction*

It is natural to investigate the brain to understand decision-making. Yet, the traditional axiomatic approach -based on introspection and developed before technological breakthroughs allowed scientists to explore neurons and brains– remains unchallenged when complex decisions and interactions are investigated.

Although several authors have made important advances in developing theoretical foundations for decision-making in the brain (e.g., Rangel et al., 2008), their contributions do not yet enable the study of games and markets in a way that compares favorably with the one enabled by classical choice theory. For example, the graduate textbook Computational Neuroscience by Gerstner et al. (2014) describes neuronal networks with electrical potential equations and time constants, but their relationship to social science concepts is obscure. A recent summary of the state of the art in Computational Neuroscience by Kass et al (2018) focuses on complementarities between mathematical and statistical approaches but the models surveyed are still far from social science.

Nevertheless, adding a neurological microfoundation to the study of decision making would allow for better predictions because preferences arise from a biological substrate. Just as Lucas (1976) argues the naivete of trying to predict the effects of a change in economic policy entirely on the basis of relationships observed in highly aggregated historical data instead of using structural parameters, it is also unwise to try to predict decision making without the structural parameters of a biology-based model of the brain. As reviewed in Bossaerts and Murawski (2015), researchers can predict a change in the preference structure in terms of structural parameters such as neurotransmitters, genetic variation and the properties of the central nervous system.

This paper provides a neural microfoundation to choice theory. The resulting "rule" can be easily applied to real-world problems, provides novel theoretical interpretations and predicts behavior that differs from that predicted by classical choice theory. The paper directly simulates neurons, aggregates them into a network known as mutual inhibition (MI), and adds refinements based on the evolutionary concepts of natural selection.

This study begins by introducing the reader to a well-established set of equations that describe the behavior of neurons. Next, it is shown that the firing frequency of a neuron when a train of incoming action potentials produces currents in the neuron can be described by a "frequency–current curve." The leaky integrate-and-fire model is the workhorse of computational neuroscience, and this paper contributes a version of it with a frequency-current curve that is simple enough to allow the next stages of modeling. Then, the paper combines distinct groups of neurons into a MI network - proven to be involved in decision-making by small animals - that contains excitatory and inhibitory channels. These behaviors are well-established characteristics of neurons. In this setting, the paper demonstrates in what circumstances this framework leads to different decision-making rules in steady states, some of which are a maximization for the organism.

A key further step is to apply the evolutionary principle of natural selection to prefer a central case among possible parameters in the MI circuit. In this central case, the steady state of the neural network can be characterized as a maximization that competes with inaction. This finding has important behavioral implications. For example, it implies that the cardinality and independence of irrelevant alternatives, which are present in classical and random utility theories, are violated in specific ways.

Another aim of this paper is to express the decision-making rule based on neuroscience in the language of the social sciences. This helps the decision rule proposed here to become simple enough to model markets and games as is customary in the social sciences. Links to Cognitive Psychology, specifically to the Drift Diffusion Model, are identified.

The remainder of the paper is structured as follows. Section 2 summarizes the evidence from neuroscience that uncovers the behaviors of single neurons and neural networks, which the subsequent model must be consistent with. Section 3 presents the model for a single neuron, derives its frequency–current curve, models the behavior of neurons acting in layers in a MI circuit, and applies refinements based on natural selection. Section 4 relabels the primary result to make it useful to social scientists. It also explores time responses and the implications of incorporating some randomness into the model. Section 5 offers observations on methodology and recommendations for future research.

## 2. The neuroscientific evidence to be captured

This section reviews some neuroscientific evidence that any behavioral model built up from a single neuron, such as the model presented in this paper, must be compatible with. This section provides essential context for the audience of social scientists.

Neuroscience and neurobiology explore brain neurons' biological structure and functions, how these building blocks are arranged to form different anatomical systems, and the neurobiology of behavior (Bear et al., 2016). Theoretical neuroscience is concerned with the development of mathematical and computational brain models that characterize what nervous systems do and how they function (Dayan & Abbot, 2005; Gerstner et al., 2014; Wang, 2012; Vogels et al., 2005). Computational neuroscience is the branch of neuroscience that models biological neural networks using computational and mathematical techniques (Arbib, 2002).

When the membrane potential of the neuron reaches a threshold, it sends a pulse to other neurons through connections called dendrites. Integrate-and-fire models of neurons are sets of equations that describe the evolution of the membrane potential over time due to incoming pulses. Integrate-and-fire models have become widely accepted as canonical models for studying both individual neurons and neural networks in Computational Neuroscience (see reviews I and II by Burkitt (2006) and the reviews by Shikri 2003 and Kass et al 2018).

Evidence shows brains (and neurons) can become exhausted. *In vivo* measurement of neurons shows repetitive and prolonged firing increases the threshold, a phenomenon called threshold fatigue (Chacron, Lindner & Longtin, A, 2007). In turn, experiments in humans suggest prolonged mental effort generates a more idling brain state and that internal attention processes are facilitated to the detriment of more extrinsic processes (Esposito et al, 2014).

Experimental evidence that neurons have delays in their behavior is substantial. Memory retrieval has been shown to be a slow process by neuroscience and psychology (see, e.g., Carrier & Pashler, 1995; Schall, 2003). Such delays are directly acknowledged by our neuron model first, and then by our network model.

Multiple and distinct neural networks that evaluate decisions in parallel are strongly supported by neuroscientific evidence (Van der Meer, 2012; Doya, 1999; Daw et al., 2005). Examples are automatic versus deliberative strategies, and selfish versus prosocial behaviors (Kiyonari et al., 2000; Rosch et al., 2000; Abbate et al., 2013; Rand et al., 2012; Rand et al., 2014; Yamagishi et al., 2007). The dorsolateral prefrontal cortex (DLPFC) and the amygdala are two distinct human brain regions involved in parallel processing. The DLPFC has been shown to regulate deliberative behaviors, inference, and reasoning (Barraclough et al., 2004; Pan et al., 2014; Donoso et al., 2014). In contrast, the amygdala has been implicated in the control of automatic behaviors such as innate response expression (Amaral, 2013), the acquisition of conditioned reactions to biologically significant stimuli (LeDoux, 1996), and automatic social decision processes (Adolphs, 2009; Haruno & Frith, 2010; Olsson & Phelps, 2007).

Evidence that brains conceive new consequences and remember past experiences provide another guide for a model like ours (Buckner, 2010). Brains also store prescribed action guides (Bunge, 2004; Morris et al., 2006). It has also been demonstrated that brains feel pleasure physically and measurably (Lecknes & Tracey, 2008). Before possible rewards are realized, they are played out in the brain (Samejima et al., 2005). A calculation similar to expected utility is performed in the brain, which is demonstrated to be a physical calculation, not an "as if" calculation (Plassmann et al., 2007; Kable & Glimcher, 2007; Knutson et al., 2005; Symmonds et al., 2010; Camille et al., 2011; Kang et al., 2011).

Any model must comprehend neural networks. Frequency–current representations of a single neuron are a well-established method. Network models can include both excitatory and inhibitory signals, the existence and differences of which are well documented. When neurons inhibit one another, competitions or races between neural groups occur. MI results in a "winner-takes-all" competition in which only one neuron group fires pulses toward the layer of neurons that implement actions. Neuroimaging, neurophysiological, and anatomical results suggest that comparisons between reward values of choice options are implemented through MI (Strait et al., 2014; Hunt et al., 2012, 2013; Jocham et al., 2012; Basten et al., 2010; Boorman et al., 2009; FitzGerald et al., 2009). Some of these competitions may take the form of a maximization (Roitman & Shadlen, 2002). Our model brings this important result from Neuroscience into sharp focus.

The evidence also suggests that parallel neural networks compete for action implementation. For example, Von Reyn et al. (2014) discovered a bimodal behavior in the genetically tractable organism Drosophila Melanogaster, a fly. When confronted with a predator-like looming stimulus, the fly responds with either a long-duration escape behavior or with a distinct, short-duration sequence that sacrifices flight stability for speed. According to recordings made during the escape, the timing of action-potentials firing in specific circuits relative to other circuits determines which of the two behavioral responses is elicited. Furthermore, Jovanic et al. (2016) identified a combination of inhibitory interneurons and disinhibition mechanisms that consolidate and maintain the chosen behavior in this fly.

Communication across a network takes time. The evidence shows that different brain areas are involved and that they activate at different times (Battels & Zeki, 2005). These steps use metabolic energy (Gershman & Wilson, 2010) and are influenced by time constraints (Ordonez & Benson, 1997). Moreover, previously chosen decisions take time to be implemented (Einhauser et al., 2010), and complexity increases the time needed for the brain to respond (see Proctor and Schneider (2018) for a review on Hick's law). These characteristics are directly embedded in our single-neuron model and appear in the organism's behavior rule.

## *3. Neuronal model of behavior*

This section aims to demonstrate how a simple description of a single neuron's physical behavior, which is both tractable and compatible with a large portion of the extensive laboratory evidence, allows us to aggregate neurons and model decision-making.

### *3.1. A simplified framework: current, voltage and frequency*

In leaky integrate-and-fire models, the most basic mathematical description in neuroscience is that neurons send and receive action potentials through connections (axons and dendrites) and direct sensory input.[2] A neuron responds by increasing or decreasing its membrane potential and sending a pulse (or action potential) to other neurons; the membrane potential at the receiving neuron adds up or "integrates" over time the currents produced by the incoming potentials. When the neuron's cumulative membrane potential reaches an upper threshold (observable, not hypothetical), the neuron issues an action potential, as shown in Figure 1:

**Figure 1: A possible time path for the membrane potential inside a given neuron**

---

[2] The internal structure of a neuron and the spatial structure of the neuron associated with the dendrites are omitted in most integrate-and-fire models. The process through which a real neuron goes through to transfer information includes filling synaptic vesicles with neurotransmitters, which then pass through a presynaptic membrane and cause the opening of ion-gated channels that change the relative permeability of the neuron.

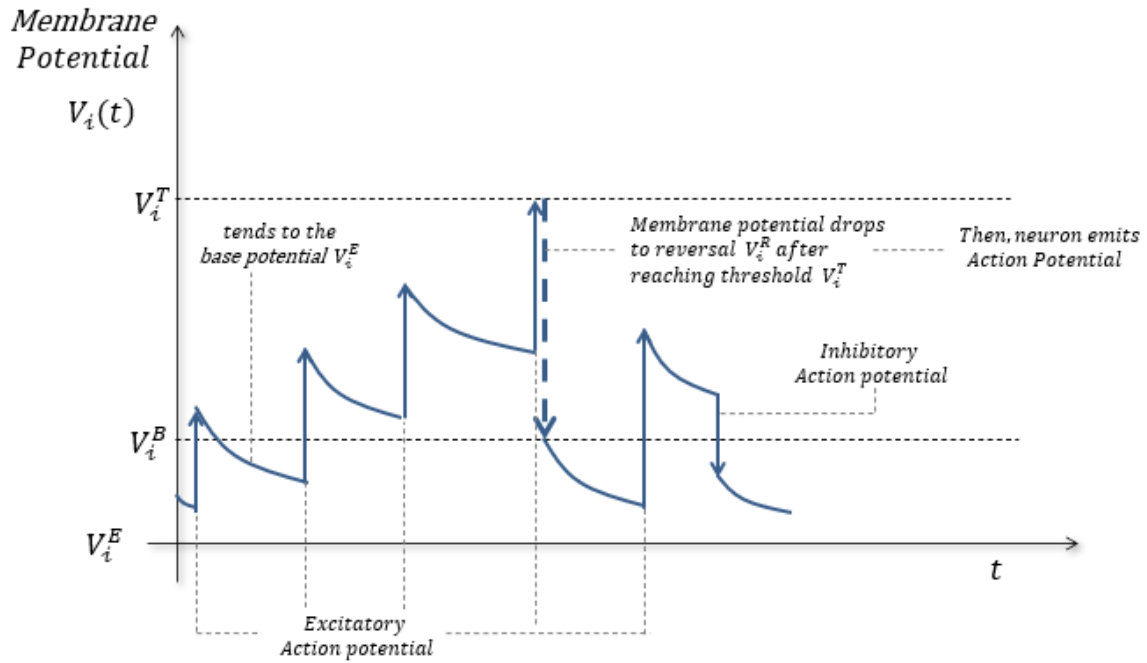

After some exogenously determined initial time, the model characterizes a neuron in terms of physical variables linked by the following equations[3]:

$$C_i \cdot \frac{dV_i(t)}{dt} = I_i(t) - C_i \cdot (V_i^T - V_i^B) \cdot \delta(V_i(t) = V_i^T) - g_i \cdot (V_i(t) - V_i^E) \quad (1i)$$

$$I_i(t) \equiv I_i^{ext}(t) + \sum_{\hbar} a_{\hbar i} \cdot q_{\hbar i}^{in}(t) \quad (1ii)$$

$$q_{\hbar i}^{in}(t) = \begin{cases} \hat{q}_{\hbar i} \cdot \delta(t) & \text{if } V_\hbar(t - \varepsilon_{\hbar i}) = V_\hbar^T \\ 0 & \text{else} \end{cases} \quad (1iii)$$

where each neuron $i$ is defined at moment $t$ by the following parameters and variables:

- The membrane potential in neuron $i$ at time $t$, $V_i(t) \in \mathbb{R}^+$, measured with respect to the base potential, $V_i^E$.
- The electrical capacitance[4] of the neuron, given by the constant $C_i \in \mathbb{R}^+$.

---

[3] The more general equations are due to Hodgkin and Huxley (1952), who built on Lapique (1907).
[4] The capacitance of an object is defined as the ratio of current to the change over time in electrical potential. Equivalently, capacitance is the ratio of the change of electric charge to the change in electrical potential.

- The current observed at neuron $i$ at time $t$, $I_i(t) \in \mathbb{R}^+$. This current affects membrane potential and is driven by incoming information, as described by (1b) and (1c).

- The upper threshold, ceiling, or maximum value for the membrane potential, at which neuron $i$ issues an action potential, $V_i^T \in \mathbb{R}^+$ towards the next layer of neurons $j$.

- Simultaneously, there is a process that reduces potential smoothly over time, called "leak" in the Neuroscience literature. Its magnitude is given by the conductance (the reciprocal of resistance) of neuron $i$'s surface, denoted by the constant $g_i \in \mathbb{R}^+$.

In equation (1i), current $I_i(t)$ is accumulated over time or "integrated" because it influences the rate of change of potential rather than its level. Equation (1i) posits that when the membrane potential in neuron $i$ reaches its maximum value $V_i^T$, it drops instantly to a reversal potential $V_i^B \in \mathbb{R}^+$, and the neuron emits an electrical potential, dubbed "action potential" by neuroscience, which is measured in the lab and labeled $q_{ij}^{out}(t)$ here, toward the next layer of neurons (it "fires").[5] A Dirac delta function $\delta(\cdot)$ is used to simplify the description of this firing. The action potential issued by one neuron is received by the next after a lapse of time ($q_{ij}^{in}(t) = q_{ij}^{out}(t - \varepsilon_{ij})$). The membrane potential in neuron $\hbar$ fires when it reaches its threshold $V_\hbar^T$. The outgoing action potential is given by $\hat{q}_{\hbar i} \cdot \delta(t) \in \mathbb{R}$, where $\hat{q}_{\hbar i}$ measures the pulse size. The time required for action potentials to travel from neuron $\hbar$ to neuron $i$ is fixed at $\varepsilon_{\hbar i} \in \mathbb{R}^+$. This delay varies across neurons $\hbar$ to capture different distances.

Because action potentials are fired through step functions in this simple model, incoming action potentials are also fired through step functions. This explains why, in Figure 1, the membrane potential jumps up when an excitatory action potential arrives, rather than increasing smoothly over time. Subscripts indicate the identity of the emitting neuron first and then the identity of the receiving neuron.

---

[5] In the lab, an action potential is a roughly 100 mV fluctuation in the potential across the neuron membrane. It lasts for about 1 millisecond (Abbott and Dayan, 2005).

If no action potentials arrive at $i$, its membrane potential decreases over time at rate $(g/C)$ and tends to an asymptotic value designated as the base potential, $V_i^E \epsilon \mathbb{R}^+$,, which is the lowest possible value of $V_i$.[6] In general, $V_i^E < V_i^B < V_i^T$.

Definition (1ii) details the following sources and types of total incoming current, $I_i(t)$:

- $I_i^{ext}(t)$ is the external current injected to neuron $i$ at time $t$ by means other than action potentials of other neurons, such as sensory organ currents.
- Neurons $\hbar$ in the preceding layer send action potentials that arrive as $q_{\hbar i}^{in}(t)$ to neuron $i$, and produce currents there.
- The magnitudes of the latter currents are influenced by specific conductances (reciprocals of resistances) labeled $a_{\hbar i} \epsilon \mathbb{R}^+$.

The following results are grounded on a simplification: first-arrival times are uniformly distributed over the duration of the accumulation phase within a potential membrane cycle. This enables us to approximate previous behavior in terms of frequency or the number of times a neuron fires in a given time interval.

Theorems 1, 2, and 3 in Appendix A shows that for the specific neural model (1$i$)-(1$ii$)-(1$iii$), the firing frequency of any one neuron $f_i(t)$ indexed by $i$ depends only on the lagged firing frequencies $f_\hbar(t - \varepsilon_{\hbar i})$ issued by neurons in the previous layer, indexed by $\hbar$ and average exterior current $\bar{I}_i^{ext}$. It can be approximated as:

$$f_i(t) \cong \alpha_i \cdot \left[\bar{I}_i^{ext}(t) + \sum_\hbar A_{\hbar i}(t) \cdot f_\hbar(t - \varepsilon_{\hbar i}) - \beta_i\right]_+ \qquad (2)$$

where $[x]_+$ denotes the maximum between 0 and $x$.[7] $\alpha_i$ and $\beta_i$ are positive constants, $\bar{I}_i^{ext}(t)$ is the exterior current coming from sensory organs and $A_{\hbar i} \equiv a_{\hbar i} \cdot \hat{q}_{\hbar i}$ is the current produced by the action-potential of cell $\hbar$ coming from neuron $i$ in the previous layer.

---

[6] This empirical fact is called "leakage" in the Neuroscience literature.

In the very short run (the milliseconds it takes a neuron to fire) the threshold $V_i^T$ remains constant, implying that $\beta_i$ is constant over time. However, the threshold fatigue effect (when sustained activity increases the firing threshold) reviewed in section 2 (Chacron et al 2007, Esposito et al 2014) tells us that over longer periods the value of $\beta_i$ rises with fatigue, as in the example in the introduction.

***Remark 1***: The signs of the currents $A_{\hbar i}(t)$ can be either positive or negative, depending on whether the connection between neuron $\hbar$ and neuron $i$ is excitatory or inhibitory. Laboratory evidence requires that inhibitory pulses be allowed.

Equation (2) states that the frequency with which neuron I fires is a (non-linear because of $[\ ]_+$) function of a linear combination of the frequencies fired into it by the previous layer of neurons. Depending on whether the connection is excitatory or inhibitory, any specific $A_{\hbar i}$ is either positive or negative. The constants $\alpha_i$ and $\beta_i$ can be expressed in terms of neuronal electrical parameters. These expressions for $\alpha_i$ and $\beta_i$ provide physical explanations for some of the results in section 4.

---

[7] In the Artificial Neuron Network literature, the function $[\ ]_+$ is called a "rectifier". If $x$ is a linear function of some other variable, the overall function is known as a "ramp function" of the other variable.

Working in the frequency domain simplifies understanding the behavior of a neural network, which is later interpreted as a choice mechanism.

## *3.2. Mutual Inhibition: a choice mechanism*

What follows focuses on a specific arrangement of neurons known as Mutual Inhibition (MI)[8], which has been linked to the act of choosing between alternatives in the neuroscience literature. As discussed in Section 2, evidence suggests that the choice behavior of small animals can be explained in terms of action-potentials in networks rich in inhibitory interneurons. The next section uses (2) to model this specific mechanism.

A neuron $i$ that is part of the MI mechanism $\mathcal{M}$ registers two different types of input signals:

(a) Action potentials $\hbar \notin \mathcal{M}$ originating outside the MI mechanism.
(b) Action potentials coming from rival neurons $j \neq i, j \in \mathcal{M}$ that are also part of the MI mechanism. The currents produced by the action potentials sent by rival neurons are always negative (inhibitory).[9]

Applying (2), the firing frequency of any one neuron $i \in \mathcal{M}$ can be written as:

$$f_i(t) = \left[ U_i(t) - \sum_{j \neq i, j \in \mathcal{M}} k_{ji} \cdot f_j(t - \varepsilon_{\hbar i}) \right]_+ \qquad (3)$$

---

[8] For an introductory text on the mathematical modeling of this mechanism, see (Gerstner et al, 2014).
[9] The mechanism works through highly responsive cells called interneurons.

Where $U_i(t)$, is defined as the uninhibited firing frequency of cell $i \in \mathcal{M}$, given by:

$$U_i(t) \equiv \alpha_i \cdot \left( \bar{I}_i^{ext}(t) + \sum_{\hbar \notin \mathcal{M}} A_{\hbar i}(t) \cdot f_\hbar(t - \varepsilon_{\hbar i}) - \beta_i \right) \quad (4)$$

Equation (4) is, of course, equation (2) with no inhibition and no threshold, i.e., the neuron's output frequency in the MI mechanism if all rivals stopped firing before applying $[\ ]_+$. As will become clear in the next section, the uninhibited firing frequency $U_i$ can be interpreted as a generalization of classical utility.

Also, $k_{ji}$ is defined as:

$$k_{ji} \equiv \alpha_i \cdot |A_{ji}(t)| \quad (5)$$

The absolute value | | and the negative sign in equation (3) are added to acknowledge currents are inhibitory. As shown in Appendix B, the $k_{ji}$ capture the relative strength of inhibitory currents produced by action potentials issued by rival neurons, by using the maximum drop in potential in the same receiving neuron as the standard. This throwback to the single neuron model confirms that modeling the behavior of individual neurons is required to understand and interpret the behavior of the network. The magnitudes of the relative inhibition ratios do not constrain the magnitude of other incoming currents.

The set of equations in (3) describes the evolution of the firing frequencies of MI neurons over time. In what follows, we assume $f_i(t)$ and $U_i(t)$ remain constant. Thus, the following system of equations is analized:

$$f_i = \left[ U_i - \sum_{j \neq i, j \in \mathcal{M}} k_{ji} \cdot f_j \right]_+ \qquad (6)$$

Characterization of the solutions to (6) will help identify the set of relative inhibitions most favored by natural selection.

**Theorem 4**: A solution to (6) always exists.

**Proof**: Use Brouwer's fixed-point theorem. See Appendix C.

Theorem 4 can be interpreted to mean that if a brain operates in a constant environment, the MI network will eventually reach a constant firing pattern, i.e. will reach a steady state.

**Theorem 5**: If there is at least one $k_{ij} > 1$ then it is always possible to find a set of uninhibited firing frequencies where, in the steady state of (6), the neuron with the smallest uninhibited firing frequency is the only active neuron. Also, this steady state lacks responsiveness to changes in the largest uninhibited firing frequency according to the set of equations in (3).

**Proof:** See Appendix C. It is sufficient to find an example.

**Theorem 6**: If a MI network is such that there is one $k_{ji} < 1$ if and $k_{ji} \cdot k_{ij} < 1$, then a set of uninhibited firing frequencies $U_i, U_j$, with a steady state where several neurons fire in parallel can always be found.

**Proof:** See Appendix C. It is sufficient to find an example.

The behaviors described in Theorems 5 and 6 may pose a threat to the organism's survival. Indeed, Theorem 5 shows that when relative inhibition parameters are larger than 1, environmental changes may not affect the organism's behavior. As the relative inhibition parameter grows (and is larger than one), the organism becomes increasingly non-responsive. Excessive non-responsiveness is detrimental to the organism's survival -in this setting- because the point of having a decision-making mechanism is to adapt towards changes in the environment.

In turn, Theorem 6 shows that multiple neurons can be active in a steady state when relative inhibition parameters are less than 1. That is, if an organism with relative inhibition smaller than one is presented with two conflicting alternatives, it may choose both alternatives concurrently, resulting in risky behavior in terms of survival, such as inconsistency, paralysis or even epileptic episodes.

A choice mechanism that selects better-adapted actions could serve as an evolutionary advantage when inheritable. Evidence suggests the brain is subject to evolutionary forces. In some young organisms, the neuron networks involved in learning expand in response to the environment over the development phase (Quartz & Sejnowski, 1997). Authors such as Adams (1998) and Fernando and Szathmary (2010) have proposed that neuron networks have a replicability feature distinct from genes, becoming units of evolution themselves and allowing brains to evolve and adapt like large ecosystems. Watson and Szathmary (2016) identify additional links between learning and evolution.

In summary, since evidence suggests the brain is molded by evolutionary dynamics, and organisms with relative inhibition parameters both greater and smaller than one face steep survival challenges, the principle of natural selection suggests that a "central case" ($k_{ji} = 1 \ \forall ji$) will become prevalent. This is the refinement based on natural selection.

For parameters $k_{ji} = 1 \ \forall ji$, Theorem 7 identifies the outgoing frequency of each MI neuron in the steady state.

**Theorem 7:** If the inhibition ratios are all equal to one ($k_{ji} = 1 \ \forall ji$) and if the list of uninhibited firing frequencies $U_i$ has a single maximum (no ties), then the steady state is unique. Moreover, the frequency issued by each MI neuron $i \in \mathcal{M}$ in that steady state is as follows:

(a) If $\max_i U_i \leq 0$, then the firing frequency of all neurons is zero. Else,

(b) The only active neuron is the one with the maximum uninhibited firing frequency $U^* = \max_{i \in \mathcal{M}} U_i$.

**Proof**: See Appendix C. The proof is by contradiction.

This means that in the "central case" identified by our refinement and described by Theorem 7, the steady state of the MI mechanism operates by maximizing across uninhibited firing frequencies, subject to competition with a positive threshold. If the threshold is not met the organism remains inactive. When the conditions for Theorem 7 are met, the MI mechanism is such that only one neuron is active in the steady state and involves a maximization. No axioms are involved in this result.

*Remark 2*: The threshold is the key that underlies the emergence of a maximization in competition with inertia. This competition distinguishes this result (Theorem 7) from a standard maximization principle in that it arises from the model and it incorporates inaction as an alternative. The existence of a threshold is due to "leakage" in equation (1i), i.e. the inevitable loss of voltage in the organism's neurons. It is also present in equation (4) in the form of a positive $\beta_i$.

In the central case, Theorem 7 implies the following behavioral rule for the MI mechanism as a whole:

$$f_\mathcal{M} = [U^*]_+ = \max(0, U^*) \; where$$
$$U^* \equiv \max_{i \in \mathcal{M}} \alpha_i \cdot \left( \bar{I}_i^{ext} + \sum_{\hbar \notin \mathcal{M}} A_{\hbar i} \cdot f_\hbar - \beta_i \right) \quad (7)$$

Where $f_\mathcal{M}$ is the outward firing frequency of the MI mechanism. The organism remains inactive when the maximum current is unable to surpass any firing threshold $\beta_i$.

***Remark 3***: Assume each neuron in the MI mechanism triggers an individual action. Then, an MI mechanism composed of N neurons allows the organism to coordinate N conflicting actions simultaneously.

## *4. Relabeling and Three Applications*

The main goal of this paper is to provide a neuroscientific microfoundation to social sciences. First, this section expresses the decision-making rule in the language of the social sciences. Next, the classical social sciences framework is adjusted to assure coherence with the neuroscientific model shown in this paper. Subsequently, a simple application of our model to response times is developed, so it can be compared with the ones proposed by Cognitive Psychology. The model is also applied to random choice, where it predicts specific deviations from ordinality and independence of irrelevant alternatives.

## *4.1. Relabeling*

In this subsection we show that the uninhibited firing frequency $U_i(t)$ in (4) can be treated as the expected utility formula in classical economic theory with some modifications. To do so, we build on Rangel et al.'s (2008) decision-making cycle, a standard decision-making framework in Neuroscience that claims favorable empirical evidence.[10] According to Rangel et al.'s (2008) cycle, choice can be broken down into several stages. Choice necessitates the brain creating a representation of the problem, reviewing possible outcomes, and selecting an action among alternatives. The MI mechanism is responsible for action selection. Moreover, Rangel et al.'s (2008) cycle posits that the action-selection stage does not receive direct pulses from the exterior ($\bar{I}_i^{ext}(t) = 0$). It is not directly connected to sensory organs, only from the stage where the problem is reviewed (labeled $\mathcal{R}$). This allows us to relabel equation (4) as neuronal utility:

$$U_i(t) \equiv \theta_i(t) \cdot \sum_{\hbar \in \mathcal{R}} \pi_{\hbar i}(t) \cdot u_{\hbar i}(t - \varepsilon_{\hbar i}) - \varphi_i \qquad (8)$$

The new terms are defined as follows, and explained below:

$$u_\hbar(t - \varepsilon_{\hbar i}) \equiv f_\hbar(t - \varepsilon_{\hbar i}) \cdot sign(A_{\hbar i}(t)) \qquad (8a)$$

$$\pi_{\hbar i}(t) \equiv \frac{|A_{\hbar i}(t)|}{\sum_{\hbar \in \mathcal{R}} |A_{\hbar i}(t)|} \qquad (8b)$$

$$\theta_i(t) \equiv \alpha_i \cdot \sum_{\hbar \in \mathcal{R}} |A_{\hbar i}(t)| \qquad (8c)$$

$$\varphi_i \equiv \alpha_i \cdot \beta_i > 0 \qquad (8d)$$

---

[10] Cisek (2007) offers an alternative to Rangel et al's five-stage cycle.

The $u_{ℏi}$ defined in (8a) has the standard classical choice model interpretation as a utility factor. $u_{ℏi}$ is defined as the product between the firing frequency function produced at the review stage, $f_ℏ$, which is always positive, and an identifier of the excitatory or inhibitory nature of the current induced in the MI mechanism, $(sign(A_{ℏi}(t)))$. This interpretation allows for negative utility levels, which are common in classical choice theory.

Interestingly, this relabeling implies that utility is a network property rather than an individual neuron property: it involves the frequency of neurons in the review stage and the excitatory/inhibitory nature of neurons in the MI mechanism. It cannot be calculated in terms of frequency of neurons in the MI mechanism alone, because the sign of the induced current would be missed.

By construction the $π_{ℏi}$ serve as subjective probabilities. According to equation (8b), these quantities depend on the relative strength of the induced currents, $|A_{ℏi}(t)|$.

Next, $θ_i$ as defined by (8c), indicates how to weigh the expected utilities relative to $φ_i$. This weight may vary with each neuron $i$. The traditional expected utility formula would link each neuron to a specific action. In that case, (8c) implies that certain actions may be preferred over others simply because of their nature.

In (8d), $φ_i > 0$ indicates the cost of computing an alternative. This cost is positive (never zero) because leak conductance $g_i$ is positive and because $V_i^E < V_i^B < V_i^T$. Also recall from section 2 that the threshold $β_i$ – and therefore $φ_i$ - increases when the brain is exhausted.

## 4.2. Adjustment to the classical social sciences framework

The standard decision-making conceptualization used in the social sciences when modelling interactions such as games and markets is Savage's (1954) analytical framework. It is characterized by the sets of acts, states of the world, consequences and a preference structure organized through axioms. This paper is focused on the mechanism that implements the comparison between choice options in a brain, and does not cover how possible acts, states of the world or consequences emerge. Nevertheless, this paper's findings help identify the simplest adjustment to the expected utility formula for it to be coherent with the neuroscientific evidence. That is, neuroscientific evidence requires agents to act -in a steady state- as maximizing neural utility (NU) subject to the inaction constraint, as in:

$$U^* \equiv \max_{\{a,p\}} \theta_p(a) \cdot \sum_{s \in S} \pi_p(a,s) \cdot u_p\big(c(a,s)\big) - \varphi_p(a)$$
$$\begin{cases} if \ U^* > 0 \quad then \quad (p,a) = arg\max_{p,a} U^* \\ else \quad \quad inaction \end{cases} \quad (NU)$$

where $a \in \mathcal{A}$ are actions, $s \in \mathcal{S}$ are states of the world, $c \in \mathcal{C}$ are consequences, the $u_p(\cdot)$ are utility functions, and $p \in \mathcal{P}$ are processes. The replacement of the $u_{\hbar i}$ in (8) for the utility functions $u_p(\cdot)$ that inherit their role in (NU) is justified by the so-called Universal Approximation (UA) theorem from the Artificial Neural Networks literature (Hornik et al 1989, Hornik 1991), which is based on the Weierstrass approximation theorem from Real Analysis. The UA theorem states that a neural network with at least one layer of neurons in-between the input and the output variables (these layers are called hidden neurons) can describe any continuous function $u: \mathbb{R}^N \to \mathbb{R}$, where $N$ is an arbitrarily large dimension size.[11] The UA theorem implies that a large enough (large N) network of physical neurons can support any choice behavior that can be described with a continuous function $u(\cdot)$.

Processes $p \in \mathcal{P}$ are explained as follows: In the classical conceptualization of choice, preferences are transitive and well-ordered, so that an individual's preference structure is represented by a single (up to linear transformations) utility function $u$. There are no such requirements in the context of this paper: many alternative valuation functions can coexist inside a brain. Indeed, allowing for multiple processes is not optional, if evidence is to be heeded: as documented in section 2, laboratory evidence strongly supports the existence of multiple and distinct neural networks that evaluate decisions in parallel (Van der Meer, 2012; Doya, 1999; Daw et al, 2005). We dub these parallel valuations "processes" $p \in \mathcal{P}$.

One implication is that contradictory assessments of the same outcome may coexist within a single brain. Another consequence of the multiplicity of preferences within a given brain is that the control of the organism's behavior may pass from one process to another as shocks to the environment occur, or as time passes by, so that a small change in the environment or in elapsed time can lead to a great change on the chosen action. To illustrate, a NU model taken to the data may identify those market conditions in which a substantial proportion of organisms suffer a change in their controlling decision-making valuations and switch behavior. For instance, the onset of panic behavior in a product market where quality is not easily verifiable could be modeled in conjunction with habitual behavior by a NU model. The NU formula implies that each transition between valuations follows a certain rule, yielding an "ordered inconsistency".

---

[11] When M parallel networks are used to describe outputs, then any function $u: \mathbb{R}^N \to \mathbb{R}^M$ can be described.

In the NU decision rule, both $\theta$ and $\varphi$ may vary by $p$ and contribute to the differences between multiple coexisting valuations. Following Rangel's model, one might hypothesize that natural selection, or perhaps learning, will favor organisms that assign higher weights $\theta$ and lower costs $\varphi$ to valuations that have previously identified superior actions.

In the classical choice setting, the states of the world are "a description of the world so complete that, if true and known, the consequences of every action would be known" (Arrow, 1971). However, the review stage in this study does not require completeness; the equivalent "states of the world" are a collection of alternative scenarios literally being evaluated by a specific brain – possibly imperfectly -- and thus are not a complete mapping of actual possibilities. This definition also allows for a broader definition of states of the world, in which emotions can play a role when probing alternative scenarios, as in (Lerner & Keltner, 2000; Tiedens et al., 2001; Lerner et al., 2003). Future work concerning the emergence of states of the world may lead to further adjustments to the neural utility formula.

### *4.3. Response times*

The general model (3) also considers the time required by the MI network to adapt when the context changes. This issue is known in Cognitive Psychology as the response time problem. The model also provides a neuroscientific microfoundation to models of this branch of social sciences.

Consider the situation in which two neurons compete within a MI mechanism where relative inhibitions are all equal to one, and the initial condition is a steady state in which neuron 1 has the highest uninhibited firing rate $U_1$. Now the context abruptly changes. Assume that neuron 2 has a smaller response time, so it is the first to exhibit an uninhibited firing rate $U_2 > U_1$, whereas neuron 1's uninhibited firing frequency remains constant at $U_1$. In our model, the response time $\Delta t$ required by the network to implement a new steady state in which only neuron 2 is active is given by:

$$\Delta t = \varepsilon_0 + \gamma \cdot (\varepsilon_{12} + \varepsilon_{21}) \cdot \frac{U_2}{U_2 - U_1} \qquad (9)$$

Where $\varepsilon_0$ is the time needed for neuron 2 to change its uninhibited firing frequency, $\gamma \in \left[1; 1 + \frac{U_2 - U_1}{U_2}\right)$ and $\varepsilon_{12}$ and $\varepsilon_{21}$ are the lags between neurons 1 and 2 and viceversa.

**Proof:** Appendix D.

One interpretation of (9) is related to models that postulate an organism accumulates abstract "evidence" until a certain threshold is reached (and then acts), such as the Drift Diffusion Model (Ratcliff, 1978; Ratcliff & McKoon, 2008), the Linear Ballistic Accumulator (Brown & Heathcote, 2008), and sequels. In this group of models, the exogenous abstract "evidence" and the threshold are replaced by the difference in firing rates and the new highest firing rate, both of which are endogenous as shown by (8).

Equation (9) is obtained when, despite the stimulus, neuron 1's uninhibited firing rate remains constant. One possible interpretation is that neuron 1 is part of a relatively "slow" network, in the sense that it is unable to produce an adapted new response before the transition time expires. However, given enough time, the slow process may yield a better result than the "fast" network, such as $U_{1+} > U_2$. In that case, (9) demonstrates how a fast process that is suboptimal in the long run can be dominant in the short run.

## 4.4. Random Mutual Inhibition

The addition of randomness to our model in equation (6) creates an alternative to McFadden's (1974) random utility theory. We show here that the properties of ordinality and independence of irrelevant alternatives are not satisfied, and that the deviation is predictable.

First, define $\overline{U}_i = E(U_i)$ as the expected value of the uninhibited firing frequency $U_i$, $\bar{f}_i = E(f_i)$ as the expectation of the post MI firing frequency $f_i$ and $\varepsilon_i \equiv (U_i - \overline{U}_i) - \sum_{j \neq i}(f_j - \bar{f}_j)$ as the error distribution concerning these quantities, which we assume has a logistic distribution with mean zero and scale parameter $\lambda$ [12]. Assuming the central case holds, (6) becomes:

$$\bar{f}_i = E\left(\left[\overline{U}_i - \sum_{j \neq i} \bar{f}_j + \varepsilon_i\right]_+\right) \qquad (10)$$

Moreover, because $\varepsilon_i$ has a logistic distribution (10) becomes:

$$\bar{f}_i = \frac{1}{\lambda}\ln\left(1 + e^{\lambda(\overline{U}_i - \sum_{j \neq i} \bar{f}_j)}\right) \qquad (11)$$

The set of equations given by (11) does not comply with the properties of ordinality and independence of irrelevant alternatives.

**Proof:** Appendix E.

This result is due to the existence of the threshold to overcome inaction. In turn, the threshold is a consequence of leakage (the inevitable loss of voltage by neurons). Because neurons are unable to maintain their voltage perfectly over time, their firing frequency will be zero at low voltages. A brain will not evaluate actions that are not vital enough to arouse it. Neurons must be stimulated sufficiently for alternatives to be considered.

---

[12] A logistic distribution with scale parameter $\lambda$ has a variance of size $\frac{\pi}{3}\lambda^2$.

***Remark 4***: in this model a negative average uninhibited firing frequency can still achieve to win the MI competition because of the error distribution used. That is, the expectation of the maximum is always strictly positive, $g_i > 0 \; \forall i$ -independently of the mean- because the logistic distribution has an infinite support.

When considering possible applications, the probability $P_i \equiv \frac{\bar{f}_i}{\sum_j \bar{f}_j}$ of neuron $i$ being active may be useful.

When the maximum incoming firing frequency $\bar{U}^* = \max_i \bar{U}_i$ is positive the result can be approximated by:

$$P_i \cong \frac{\ln(1 - e^{-\lambda(U^* - \bar{U}_i)} e^{-\omega})}{\sum \ln(1 - e^{-\lambda(U^* - \bar{U}_j)} e^{-\omega})} \qquad (12)$$

Where $\omega = \frac{e^{-\lambda \bar{U}^*}}{\prod_{j \neq i^*}\left(1 - e^{-\lambda(\bar{U}^* - \bar{U}_j)}\right)} > 0$, and $i^*$ is the neuron perceiving $\bar{U}^*$.

When the maximum incoming firing frequency $\bar{U}^* = \max_i \bar{U}_i$ is negative the result can be approximated by:

$$P_i \cong \frac{\ln(1 - e^{\lambda \bar{U}_i})}{\sum \ln(1 - e^{\lambda \bar{U}_j})} \qquad (13)$$

It is also useful to note that despite choice being discrete, a continuous statistic (the time needed to reach a decision) carries information regarding the strength of the winning alternative. In particular, the time needed to reach a decision is the sum of the time needed to create a description of the problem at hand and the inverse of the frequency needed to reach that decision (the period, in units of time). The expected value of the period (inverse of the frequency) can be approximated, simulated or calculated by other means in terms of the already found quantities above.

The previous conceptualization allows us to prove Hick's law, which relates reaction time to the logarithm of the number of stimulus-response alternatives. That is, our model implies:

$$\frac{dT}{dN} = \frac{1}{\bar{f}_i} \frac{\partial}{\partial N} \ln\left(e^{-\lambda \bar{f}_i} + \left(1 - e^{-\lambda \bar{f}_i}\right) N\right) \quad (14)$$

**Proof:** Appendix F.

Summarizing section 4, the relabeling of uninhibited firing frequency allows rewriting equations (8), (9), and (11) in more traditional terms. According to (8), in stable environments an organism will maximize utility, but only if it overcomes inaction. Equation (9) shows that even if the organism can maximize utility in the long run, decisions take time and fast responses may be prevalent in the short run. Meanwhile, equation (11) indicates that utility does more than just order preferences (ordinality): because inaction needs to be overcome, the level of utility is important when predicting outcomes (cardinality). Because the inaction threshold prevents pure maximization, the property of independence of irrelevant alternatives is satisfied less often as maximized utility comes closer to this threshold.

## *5. Concluding remarks*

This study exemplifies a new general method for obtaining decision-making rules based on neuroscience models. In this specific paper, we begin with a simplified model for a single neuron, add aggregation in the frequency domain, and analyze a specific brain circuitry whose steady states can be characterized using fixed point theorems. Later, the natural selection principle is used as a refinement to select prevalent network parameters that lead to a behavioral rule.

This general method has major methodological differences with classical justifications of decision-making rules. One such justification is that individual organisms have the intention to do better, given preferences. In contrast, the decision rule found here, which also includes a maximization, does not result from an intentional entity. In science, unintended maximizations are not uncommon: for example Arrow and Debreu (1954) showed that an economy without externalities and asymmetric information, in which each agent independently chooses her consumption and production bundles while accepting prices as given, yields the same allocation of resources as if a single mind actively organized society. The specific mechanisms derived here are consistent with much of the hard evidence neuroscience has accumulated over the past 30 years, as detailed in section 2. In the extension which we designate as Neuronal Utility (NU), multiple processes allow for systematic shifts in the valuation process that control actions and implementation, as required by the evidence. A suggested avenue for future research in neuroeconomics is to overcome the main limitation of this study, which centers on completing a thorough justification for transiting from neuroscience-based rules such as (7) towards easily applied decision rules such as (NU).

Specifically, actions, states of the world and consequences still need to be associated with patterns observable by neuroscience.

The model provides a coherent neuroscientific microfoundation to both Economics Choice Theory and Cognitive Psychology Response-Time Theory.

## *Appendix A: Integrate and fire model results*

Time-averaging methods translate equations (1i-iii) from potentials in the time domain into frequencies. A focus on frequency reduces the mathematical complexity of the information-transmission process, especially for large groups of neurons (populations) and networks.

The average rate at which each neuron issues pulses, or the frequency of the outgoing pulses, is computed as follows (Ermentrout, 2008). The "firing rate" or frequency of a neuron is defined as the number of pulses issued per unit of time. For averaging to be precise, this interval must include a large number of neuronal cycles.

An outside observer who is just starting to measure would have no idea when the first pulse in a train would arrive. Our simple model makes the unique assumption that after the first pulse, the subsequent incoming pulses—and the current produced—have a 100% positive correlation over time. They are ex-post deterministic. This feature also implies that the receiving neuron can precisely predict when the next outgoing pulse will be issued.

**Theorem 1**: If the distributions of the arrival times of the first observed incoming action potentials in trains of such pulses that are 100% positively correlated over time and come from each neuron in the previous layer, are uniform over the subthreshold lapse and independent between neurons, then an approximation for the expected current over time observed at neuron $i$ within an averaging interval that lasts an adequately large number of cycles in membrane potential is obtained:

$$\bar{I}_i(t) \cong \sum_\hbar A_{\hbar i} \cdot f_\hbar(t - \varepsilon_{\hbar i}) + \bar{I}_i^{ext}(t) \qquad (A1)$$

where $A_{\hbar i} \equiv a_{\hbar i} \cdot \hat{q}_{\hbar i}$ is the current produced by the action-potential of cell $\hbar$ in neuron $i$, $f_\hbar$ is the frequency at which neuron $\hbar$ in the previous layer issues pulses towards neuron $i$, and $\bar{I}_i^{ext}$ is the average external current over an interval starting at time $t$.

**Proof**: Start with equation (1ii)

$$I_i(t) \equiv I_i^{ext}(t) + \sum_\hbar a_{\hbar i} \cdot q_{\hbar i}^{in}(t) \qquad (A1a)$$

Take expectations on both sides:

$$\bar{I}_i(t) \equiv \mathbb{E}(I_i(t)) = \mathbb{E}(I_i^{ext}(t)) + \sum_\hbar a_{\hbar i} \cdot \mathbb{E}\left(q_{\hbar i}^{in}(t)\right) \quad (A1b)$$

Define $\bar{I}_i^{ext}(t) \equiv \mathbb{E}(I_i^{ext}(t))$. To determine $\mathbb{E}\left(q_{\hbar i}^{in}(t)\right)$ assume that the distribution of the arrival time of action-potentials is uniform. Also use equation (1 iii), so action-potentials are fired only when the membrane potential reaches a threshold and the pulse takes a lapse to reach the next layers of neurons::

$$\mathbb{E}\left(q_{\hbar i}^{in}(t)\right) = \int_0^{1/f_\hbar(t-\varepsilon_{\hbar i})} \left(\hat{q}_{\hbar i}\delta(t)\right) d\mathbb{P}(V_\hbar(t-\varepsilon_{\hbar i}) = V_\hbar^T) \qquad (A1c)$$

$$\mathbb{E}\left(q_{\hbar i}^{in}(t)\right) = \hat{q}_{\hbar i} \int_0^{1/f_\hbar(t-\varepsilon_{\hbar i})} \delta(t) d\mathbb{P}(V_\hbar(t_\hbar) = V_\hbar^T) = \hat{q}_{\hbar i} \int_0^{1/f_\hbar(t-\varepsilon_{\hbar i})} \delta(t) \frac{dt}{(1/f_\hbar(t - \varepsilon_{\hbar i}))} \qquad (A1d)$$

$$\mathbb{E}\left(q^{in}_{\hbar i}(t)\right) = \hat{q}_{\hbar i} \cdot f_{\hbar}(t - \varepsilon_{\hbar i}) \int_0^{1/f_{\hbar}(t-\varepsilon_{\hbar i})} \delta(t)dt = \hat{q}_{\hbar i} \cdot f_{\hbar}(t - \varepsilon_{\hbar i}) \quad (A1e)$$

Defining $A_{\hbar i} \equiv a_{\hbar i} \cdot \hat{q}_{\hbar i}$ and collecting terms we find that:

$$\bar{I}_i(t) = \sum_{\hbar} A_{\hbar i} \cdot f_{\hbar}(t - \varepsilon_{\hbar i}) + \bar{I}^{ext}_i(t) \quad \text{QED.}$$

According to Equation (A1), the expected current at neuron $i$ is determined by a linear combination of currents produced in the previous layer of neurons. The incoming frequencies are represented by the weights. It is worth noting that a frequency of zero generates no current.

Consider the relationship between the emitted frequency and the time-average of the current observed within the given neuron. In computational Neuroscience, this relationship is known as the $f - I$ curve.

**Theorem 2:** If the arrival times of the first observed incoming action potentials from the previous layer are uniform and independent, then the firing rate of outgoing pulses is approximately:

$$f_i \cong \left[\frac{\bar{I}_i - g_i \cdot (\bar{V}_i - V^E_i)}{C_i \cdot (V^T_i - V^B_i)}\right]_+ \quad (A2)$$

where the notation $[x]_+$ indicates the maximum between 0 and $x$, and $\bar{I}_i$ is the average current over time observed at neuron $i$.

**Proof**: Analyze first the case where incoming pulses generate a current strong enough to overcome leak conductance and cause neuron $i$ to emit pulses with a frequency of $f_i$. As in Theorem 1, take expectations with respect to the moment of the first incoming pulses from the previous layer in equation (1i).

$$C_i \cdot \mathbb{E}\frac{dV_i(t)}{dt} = -g_i \cdot (\mathbb{E}[V_i(t)] - V^E_i) + \mathbb{E}[I_i(t)] - C_i \cdot (V^T_i - V^B_i) \cdot \mathbb{E}[\delta(V_i(t) = V^T_i)] \quad (A2a)$$

where the incoming pulses induce current $I_i(t)$ from those first moments onwards.

Next, define $t_0$ as the present instant, N as the number of neuron cycles within the averaging interval and $t_0 + \frac{N}{f_i}$ as the instant in the last cycle of the averaging interval (N cycles later) where the averaging interval ends. Now sum over time between $t_0$ and $t_0 + \frac{N}{f_i}$:

$$C_i \cdot \int_{t_0}^{t_0 + \frac{N}{f_i}} \mathbb{E}[dV_i(u)] =$$

$$-g_i \cdot \left(\int_{t_0}^{t_0 + \frac{N}{f_i}} \mathbb{E}[V_i(u)]du - \frac{N \cdot V^E_i}{f_i}\right) + \int_{t_0}^{t_0 + \frac{N}{f_i}} \mathbb{E}[I_i(u)]du - C_i \cdot (V^T_i - V^B_i)\int_{t_0}^{t_0 + \frac{N}{f_i}} \mathbb{E}[\delta(V_i(u) = V^T_i)]du \quad (A2b)$$

Then calculate each term:

$$\int_{t_0}^{t_0 + \frac{N}{f_i(t_0)}} \mathbb{E}[dV_i(u)] = V_i\left(t_0 + \frac{N}{f_i(t_0)}\right) - V_i(t_0) \quad (A2c)$$

$$\int_{t_0}^{t_0+\frac{N}{f_i(t_0)}} \mathbb{E}[V_i(u)]du \equiv \frac{N \cdot \bar{V}_i(t_0)}{f_i(t_0)} \qquad (A2d)$$

$$\int_{t_0}^{t_0+\frac{N}{f_i(t_0)}} \mathbb{E}[I_i(u)]du \equiv \frac{N \cdot \bar{I}_i(t_0)}{f_i(t_0)} \qquad (A2e)$$

$$\int_{t_0}^{t_0+\frac{N}{f_i(t_0)}} \delta(V_i(u) = V_i^T)du = N \qquad (A2f)$$

where $\bar{V}_i(t_0)$ and $\bar{I}_i(t_0)$ are defined as the time averages of the membrane potential and the current, during the averaging period starting at $t_0$. Equation (A2f) comes from the definition of a Dirac delta. Replacing all terms, the result is:

$$f_i(t_0) = \frac{\bar{I}_i(t_0) - g_i \cdot (\bar{V}_i(t_0) - V_i^E)}{C_i \cdot (V_i^T - V_i^B) + C_i \cdot \frac{[V_i(t_0 + N/f_i) - V_i(t_0)]}{N}} \qquad (A2g)$$

Taking a large enough number of intervals N and observing that the membrane potential is bounded, the last term in the denominator is approximately zero.

When incoming pulses generate a current that is too weak to overcome leak conductance, the frequency must be zero. This aspect necessitates the use of the $[x]_+$ function. QED.

Theorem 3 then shows that, in this particular model, the $f - I$ curve is bounded by two parallel straight lines that act as asymptotes, are time-independent, linear in current, have the same slope, and can be relatively close to each other.

**Theorem 3**: If the expected current's variation over time is bounded and Theorem 2 holds, then bounds on the outgoing frequency exist, are independent of the average membrane potential, are linear in current $\bar{I}_i$, and are given by:

$$\left[\frac{\bar{I}_i(t) - g_i \cdot (V_i^T - V_i^E)}{C_i \cdot (V_i^T - V_i^B)}\right]_+ \leq f_i(t) \leq \left[\frac{\bar{I}_i(t) - g_i \cdot (0{,}5 \cdot [V_i^T + V_i^K] - V_i^E)}{C_i \cdot (V_i^T - V_i^B)}\right]_+ \qquad (A3)$$

where the bound on the expected current is given by: $\frac{d\mathbb{E}[I_i(t)]}{dt} \leq g_i \cdot \frac{d\mathbb{E}[V_i(t)]}{dt}$.

**Proof.** Frequency can be positive only if the membrane potential rises on average over time within a cycle. This in turn requires the size of the expected current to be large enough, i.e. according to (1i), it is required that

$$C_i \cdot \frac{d\mathbb{E}[V_i(t)]}{dt} = -g_i \cdot \left(\mathbb{E}[V_i(t)] - V_i^E\right) + \mathbb{E}[I_i(t)] > 0 \ , \forall t \qquad (A3a)$$

Take the derivative of (T3) with respect to time and impose a slow enough change in current:

$$\frac{d\mathbb{E}[I_i(t)]}{dt} \leq g_i \cdot \frac{d\mathbb{E}[V_i(t)]}{dt} \qquad (A3b)$$

It follows that

$$C_i \cdot \frac{d^2\mathbb{E}[V_i(t)]}{dt^2} \leq 0, \ \forall t \qquad (A3c)$$

which implies that on average the membrane potential is concave or linear with respect to time. Concavity in turn implies that:

$$\bar{V}_i(t) \geq 0{,}5 \cdot [V_i^T + V_i^B] \quad (A3d)$$

This, plus Theorem 2 and the fact that $\bar{V}_i(t)$ is always less than the upper bound $V_i^T$, proves Theorem 3. QED

The existence of these bounds suggests an approximate version for the $f - I$ curve. Specifically, from this point forward each neuron's behavior will be represented here by the approximate function $f_i(t) \cong \alpha_i \cdot [\bar{I}_i - \beta_i]_+$, where the coefficients obtain from Theorem 3 as $\alpha_i \equiv 1/\left(C_i \cdot (V_i^T - V_i^B)\right)$ exactly and as $\beta_i \in [g_i \cdot (0{,}5 \cdot [V_i^T + V_i^B] - V_i^E);\ g_i \cdot (V_i^T - V_i^E)\ ]$.

The exact expression for $\alpha_i$, turns out to be essential in Section 4 to interpret the decision-making model and to complete it. It is also important that $\beta_i > 0$, a result that follows from the fact that conductance $g_i$ is positive and that $V_i^E < V_i^B < V_i^T$.

Combining this approximate version with Theorem 1 leads to a simple description of the behavior of a single neuron $i$ in the frequency domain:

$$f_i(t) \cong \alpha_i \cdot \left[ \bar{I}_i^{ext}(t) + \sum_{\hbar} A_{\hbar i} \cdot f_{\hbar}(t - \varepsilon_{\hbar i}) - \beta_i \right]_+ \quad (2)$$

Equation (2) is our workhorse model for single neuron behavior. (2) with time lags, connects the outgoing frequency from neuron i to the currents produced in the same neuron by the incoming action potentials from the previous layer of neurons, with time lags. This linear form with a nonlinearity at zero is called a "rectified linear activation function" in the artificial neuron network literature.

There is a wealth of neuroscientific research on actual $f - I$ curves. The $f - I$ curve for our simple model is consistent with the experimental behavior of some neurons, according to the literature (Azouz et al., 1997; Ahmed et al., 1998). However, other neurons exhibit differences, such as a decreasing current slope (Agmon & Connors, 1992; Rauch et al., 2003). Extensions that incorporate these distinctions are left to future research.

## *Appendix B: Relative inhibitory strength*

The ratio $k_{ji}$ captures the relative strength of inhibitory currents produced by actions potentials issued by rival neurons, using as measuring standard the maximum drop in potential in the same receiving neuron. Specifically:

$$k_{ji} \equiv \frac{-\frac{\Delta V_i(t)}{\Delta t}}{(V_i^T - V_i^B)} = \frac{\left(\frac{1}{C_i}\right) \cdot |A_{ji}|}{(V_i^T - V_i^B)} \quad (B)$$

The second equality uses equations (1i) and (1ii).

The expression for the relative drop in potential is simplified further by using the definition of coefficient $\alpha_i$, obtained from Theorem 3 as $\alpha_i \equiv 1/\left(C_i \cdot (V_i^T - V_i^B)\right)$. This yields the relative inhibition ratio shown in (5).

## *Appendix C: Steady State in Mutual Inhibition*

**Theorem 4**: A solution to (6) always exists.

**Proof:** Brouwer's fixed-point theorem states that for any continuous function $T$ mapping a compact convex set to itself there is at least one point $x_0$ such that $T(x_0) = x_0$.

Her, since $f_i \geq 0 \; \forall i$, the function

$$T(f_i) = \left[ U_i - \sum_{j \neq i} k_{ji} \cdot f_j \right]_+ \quad \forall i,j \quad (C1)$$

is continuous and maps a compact set (remember $f_i \in [0, [U_i]_+] \; \forall i$) into itself. Thus, a fixed point of (6) always exists. QED

**Theorem 5**: If there is at least one $k_{ij} > 1$ then it is always possible to find a set of uninhibited firing frequencies where, in the steady state of (6), the neuron with the smallest uninhibited firing frequency is the only active neuron. Also, this steady state lacks responsiveness to changes in the largest uninhibited firing frequency according to the set of equations in (3).

**Proof:** It is sufficient to find an example. Choose a set of uninhibited firing frequencies where only two neurons, $j$ and $i$, have strictly positive uninhibited firing frequencies. As a result, the other neurons are inactive and can be excluded from the analysis. Assume that the neuron with the highest uninhibited firing frequency is $j$ ($U_j > U_i$). Assume that $j$ is inactive and $i$ is active. Then

$$f_j = 0 \; ; \; U_j < k_{ij} \cdot f_i \quad (C2a)$$
$$f_i = U_i \quad (C2b)$$

Then

$$U_j < k_{ij} \cdot U_i \quad (C2c)$$

Thus, an example proving the theorem only requires finding some $U_i$ such that

$$\frac{U_j}{k_{ij}} < U_i < U_j \quad (C2d)$$

Since by assumption $k_{ij} > 1$, result (C2d) is always possible.

Now we must prove the resulting equilibrium lacks responsiveness to changes in the largest uninhibited firing frequency. Use (3) and assume -again- that only neurons $i$ and $j$ have positive uninhibited firing frequencies. Then the MI system remains constant as long as:

$$f_j(t) = 0 \; ; \; U_j(t) < k_{ij} \cdot f_i(t - \varepsilon) \quad (C2e)$$
$$f_i(t) = U_i \quad (C2f)$$

That is, when $k_{ij} > 1$ there are sets of firing frequencies where the evolution of the MI system remains constant. In the counter example shown above, as long as $U_j(t) < k_{ij} \cdot U_i$ the resulting equilibrium will remain undisturbed. That is, as $k_{ij}$ augments the system becomes increasingly non-responsive. QED.

**Theorem 6**: if there is a $k_{ij} < 1$ and $k_{ij} \cdot k_{ji} < 1$, then a set of uninhibited firing frequencies where in the steady state several neurons fire in parallel can always be found for (6).

**Proof:** It is sufficient to find an example. Select an uninhibited firing frequency set where only two neurons, $j$ and $i$, have strictly positive uninhibited firing frequencies. Therefore, the other neurons are inactive and can be excluded from the analysis. Assume the neuron with the largest uninhibited firing frequency is $j$ ($U_j > U_i$). Both $i$ and $j$ are active. Then

$$f_j = U_j - k_{ij} \cdot f_i \quad (C3a)$$
$$f_i = U_i - k_{ji} \cdot f_j \quad (C3b)$$

Then, we can find

$$f_i = \frac{U_i - k_{ji} \cdot U_j}{1 - k_{ij} \cdot k_{ji}} \quad (C3c)$$

$$f_j = \frac{U_j - k_{ij} \cdot U_i}{1 - k_{ij} \cdot k_{ji}} \quad (C3d)$$

From (C3d) it follows that $f_j > 0$ because $U_j > U_i$, $k_{ij} < 1$ and $k_{ij} \cdot k_{ji} < 1$

Now we find the conditions under which $f_i > 0$

Assuming $U_j = U_i \cdot (1 + e)$ with $e > 0$ so that $U_j > U_i$

We can find

$$f_i = \frac{U_i \cdot \left(1 - k_{ji} \cdot (1 + e)\right)}{1 - k_{ij} \cdot k_{ji}} \quad (C3e)$$

Thus, an example proving the theorem only requires $k_{ji} \cdot (1 + e) < 1$. That is, it is only needs to find a $U_i$ such that

$$U_j \cdot k_{ji} < U_i < U_j \quad (C3f)$$

This is always possible because the Theorem assumes that $k_{ij} < 1$. QED.

**Theorem 7:** If the inhibition ratios are all equal to 1 ($k_{ji} = 1 \ \forall j, i$) and the list of uninhibited firing frequencies has a single maximum (is single peaked), then the steady state is unique. Specifically,

(a) If $\max_i U_i = 0$, then all neurons exhibit Inaction. Else,
(b) Only the neuron with the largest uninhibited firing frequency is active.

This can be summarized as

$$\begin{cases} \text{if} \quad U_i = \max_j U_j \quad \text{then} \quad f_i = U_i \\ \text{else} \quad f_i = 0 \end{cases} \quad (C4)$$

**Proof:** The steady state of the network at the MI network, in this case, is determined by the following system of equations:

$$f_i = \left[ U_i - \sum_{j \neq i} f_j \right]_+ \quad (C4a)$$

Let us review each case.

(a) Inaction. If the average current for each action championed at the MI stage is unable to exceed the threshold $\beta_i$, then the firing frequency of every MI neuron is zero. The organism cannot act because no signals are sent to the implementing neurons. In other words, inaction occurs if and only if $\max_i U_i \leq 0$. Else,

(b) To demonstrate: Only the neuron with the highest uninhibited firing frequency is active.

To begin, we demonstrate by contradiction that the neuron with the highest uninhibited frequency must be active, which can only be one because no ties are allowed by assumption. Assume this neuron is dormant. Then one or more other neurons fire. The term inside the brackets $[\ ]_+$ is then positive for this group of neurons. Consider resolving the system of equations (C4a) for the active neuron subset. Each equation can be rewritten by shifting the frequencies of the active competing neurons to the left-hand side. Thus, the left side of each equation is the sum of all frequencies in the active set. Because the right-hand sides only contain the uninhibited firing rates of the various active neurons, the sum of the frequencies can be no greater than the second largest uninhibited firing frequency. When this result is entered into the equation for the neuron with the highest uninhibited frequency, its frequency must be positive, indicating that it is active. This demonstrates the contradiction.

Second, we show that only one neuron must be active by using contradiction. Assume that at least two neurons are active. As previously demonstrated, one of them must be the neuron with the highest uninhibited frequency. For this set of neurons, the function $[\ ]_+$ is always positive. Consider solving the equation system (C4a). Each equation can be rewritten by shifting the frequencies of competing neurons to the left side. As a result, the left side of each equation is the sum of all frequencies in the active set. On the other hand, the right-hand side only contains the uninhibited firing rates of the other (different) active neurons. There cannot be a tie between the active neurons because the list of uninhibited firing frequencies has a single maximum.

As a result, in this case, only one neuron is active. That neuron has the highest uninhibited frequency. QED.

## *Appendix D: Response times after a simple shock*

**Theorem 8:** Assume an initial steady state where there is only one active mutual-inhibition neuron, labeled 1, whose uninhibited firing frequency is $f_1 = U_1$. Assume that $k_{ji} = 1 \; \forall j, i$. Assume that at time $t_0$ an external stimulus changes context for the organism. Assume one of the processes adapts and a second neuron, labeled 2, champions a new action proposal.

Assume the new action proposal is competitive. That is, neuron 2 increases its uninhibited firing frequency from $f_2 = U_{2-} < U_1$ strictly to $f_2 = U_2 > U_1$ strictly. This new action proposal reaches the action selection stage at time $t_0 + \varepsilon_0$, where $\varepsilon_0$ is a sum of processing times in the previous stages.

Assume that the stimulus doesn't change neuron 1's firing frequency: its process doesn't adapt fast enough to be considered in this transition. In the MI mechanism, information is transferred with lags $\varepsilon_{12}$ and $\varepsilon_{21}$.

After some transition, neuron 2 becomes the only active neuron. Arrival to the new steady state, which will trigger a new response, takes a total time $\Delta t$ given by:

$$\Delta t = \varepsilon_0 + \gamma \cdot (\varepsilon_{21} + \varepsilon_{12}) \frac{U_2}{U_2 - U_1} \qquad (D1)$$

**Proof:** Equation (6) indicates the frequencies of neurons 1 and 2 at any time $t$. Using the assumption that $k_{12} = k_{21} = 1$:

$$f_1(t) = [U_1 - f_2(t - \varepsilon_{21})]_+ \qquad (D1a)$$
$$f_2(t) = [U_2 - f_1(t - \varepsilon_{12})]_+ \qquad (D1b)$$

with initial conditions $f_2(t_0 + \varepsilon_0) = 0$ and $f_1(t_0 + \varepsilon_0) = U_1$.

Now we take into account that a cycle of MI requires two interactions, each with its own lag, to come full circle back to neuron 2. The lags add up. In equation (D1a), change variables to $t' = t - \varepsilon_{21}$. In equation (D1b), change variables to $t'' = t - \varepsilon_{12}$.

$$f_1(t' + \varepsilon_{21}) = [U_1 - f_2(t')]_+ \qquad (D1c)$$
$$f_2(t'' + \varepsilon_{12}) = [U_2 - f_1(t'')]_+ \qquad (D1d)$$

During the transition both neurons are active. Then, function $[\ ]_+$ can be dropped.

There is a connection between the times at which the two competing neurons modify their frequencies:

$$t' = t'' - \varepsilon_{21} \qquad (D1e)$$

Replace $(D1e)$ in $(D1c)$ to get:

$$f_1(t'') = U_1 - f_2(t'' - \varepsilon_{21}) \qquad (D1f)$$

Replace $(D1f)$ in $(D1d)$, to find how the actual frequency of the second neuron rises over time, in a cumulative fashion (in parallel, the frequency of neuron 1 falls over time):

$$f_2(t'' + \varepsilon_{12}) = U_2 - \{U_1 - f_2(t'' - \varepsilon_{21})\} \qquad (D1g)$$

To simplify further, define $z \equiv t'' - \varepsilon_{21}$. Replacing:

$$f_2(z + \varepsilon_{21} + \varepsilon_{12}) - f_2(z) = U_2 - U_1 \qquad (D1h)$$

Equation $(D1h)$ says that the frequency of neuron 2 increases over time with increments of size $U_2 - U_1$ after each block of $\varepsilon_{21} + \varepsilon_{12}$ time units. These blocks can be interpreted as the time needed to complete a single cycle of MI. Since these blocks have a fixed and finite size, convergence to the new steady state occurs within a finite number of mutual inhibition cycles. When mutual inhibition is repeated $N$ times, we find

$$f_2(t_0 + \varepsilon_0 + N \cdot (\varepsilon_{21} + \varepsilon_{12})) - f_2(t_0 + \varepsilon_0) = N \cdot (U_2 - U_1) \quad (D1i)$$

Define $\Delta t$ as the total time needed by neuron 2 to reach its uninhibited firing frequency. It defines the time the network takes to reach its new steady state. Without loss of generality, assume that the time required is in between $N$ and $N + 1$ cycles of mutual inhibition. Then:

$$N \leq \frac{U_2}{U_2 - U_1} < N + 1 \quad (D1j)$$

By the mean value theorem, there exists an $\widehat{N} \in [N; N + 1)$ such that

$$\Delta t = \varepsilon_0 + \widehat{N} \cdot (\varepsilon_{21} + \varepsilon_{12}) \quad (D1k)$$

Eliminating $\widehat{N}$ with the inequations in $(D1j)$ there is $\gamma$ such that:

$$\Delta t = \varepsilon_0 + \gamma \cdot (\varepsilon_{21} + \varepsilon_{12}) \cdot \frac{U_2}{U_2 - U_1} \quad (D1l)$$

where coefficient $\gamma \in \left[1; 1 + \frac{U_2 - U_1}{U_2}\right)$ registers the slack in the last mutual inhibition cycle. QED.

Note: The number of cycles required to reduce neuron 1's frequency to zero is $U_1/(U_2 - U_1)$. This in turn is equal to $N - 1$, where $N$ is the number of mutual inhibition cycles required by neuron 2 to reach its new uninhibited frequency.

## *Appendix E: Random mutual inhibition*

**Theorem 9:** First, define $\overline{U}_i$ as the expected value of uninhibited firing frequency, $\bar{f}_i$ as the expectation of the resulting MI firing frequency of $f_i$ and $\varepsilon_i \equiv (U_i - \overline{U}_i) - \sum_{j \neq i}(f_j - \bar{f}_j)$, which distributes logistic. Assuming the central case holds, (6) becomes:

$$g_i = E\left(\left[\overline{U}_i - \sum_{j \neq i} \bar{f}_j + \varepsilon_i\right]_+\right) \quad (E1a)$$

Now remember that if $x$ is a random error that has a logistic distribution with media $\mu$ and constant scale parameter $\lambda$, then, $E[x]_+ = \frac{1}{\lambda}\ln(1 + e^{\lambda \mu})$. Therefore, (E1a) becomes:

$$\bar{f}_i = \frac{1}{\lambda}\ln\left(1 + e^{\lambda(\overline{U}_i - \sum_{j \neq i} \bar{f}_j)}\right) \quad (E1b)$$

The previous equation does not have an analytic solution, yet it can be reordered to simplify the analysis. First, solve for $\bar{f}_i$ in terms of $m = \sum \bar{f}_j$;

$$e^{\lambda \bar{f}_i} = 1 + e^{\lambda(\overline{U}_i - \sum \bar{f}_{j \neq i})} \quad (E1c)$$
$$e^{\lambda \bar{f}_i} = 1 + e^{\lambda \overline{U}_i} e^{-\lambda \sum \bar{f}_j} e^{\lambda \bar{f}_i} = 1 + e^{\lambda \overline{U}_i} e^{-\lambda m} e^{\lambda \bar{f}_i}$$
$$e^{\lambda \bar{f}_i} = \frac{1}{1 - e^{\lambda \overline{U}_i} e^{-\lambda m}} \quad (E1d)$$
$$\bar{f}_i = -\frac{1}{\lambda}\ln\left(1 - e^{-\lambda(m - \overline{U}_i)}\right) \quad (E1e)$$

Thus, a solution must, in parallel, comply with:
$$\begin{cases} m = -\dfrac{1}{\lambda}\sum \ln\left(1 - e^{-\lambda(m-\bar{U}_j)}\right) \\ \bar{f}_i = -\dfrac{1}{\lambda}\ln\left(1 - e^{-\lambda(m-\bar{U}_i)}\right) \end{cases} \quad (E1f)$$

Now, if we define $P_i \equiv \dfrac{\bar{f}_i}{\sum \bar{f}_j}$ then:
$$P_i = \frac{\ln\left(1 - e^{-\lambda(m-\bar{U}_i)}\right)}{\sum \ln\left(1 - e^{-\lambda(m-\bar{U}_j)}\right)} \quad (E1g)$$

This formula is different to McFadden's (1974) random utility theory and does not meet the properties of Ordinality and of Independence of Irrelevant Alternatives.

Ordinality:
Let us add a constant $\theta \neq 0$ to all uninhibited firing frequencies:
$$\widetilde{U}_i = \bar{U}_i + \theta \quad \forall i \quad (E1h)$$
Then $\widetilde{m}$ should become $m + \theta$ so that $P_i$ remains constant for all $\lambda$ in (E1g). Nevertheless, by definition
$$\widetilde{m} = -\frac{1}{\lambda}\sum \ln\left(1 - e^{\lambda \widetilde{U}_j} e^{-\lambda \widetilde{m}}\right) = -\frac{1}{\lambda}\sum \ln\left(1 - e^{\lambda \bar{U}_j} e^{-\lambda m}\right) = m \quad (E1i)$$
Which is a contradiction because we assumed $\theta \neq 0$. The ordinality property is not met because negative uninhibited firing frequencies do not generate responses in the MI setting.

Independence of irrelevant alternatives:
Let us take the ratio between the probabilities of alternatives $i$ and $j$:
$$\frac{P_i}{P_j} = \frac{\ln\left(1 - e^{-\lambda(m-\bar{U}_i)}\right)}{\ln\left(1 - e^{-\lambda(m-\bar{U}_j)}\right)} \quad (E1j)$$

This ratio is not independent of the uninhibited firing frequencies of other alternatives because m cannot be eliminated in the equation, an m ultimately is a function of all uninhibited firing frequencies of all alternatives. In fact:
$$\frac{dm}{d\bar{U}_h} = \frac{\dfrac{1}{e^{\lambda(m-\bar{U}_h)} - 1}}{1 + \sum \dfrac{1}{e^{\lambda(m-\bar{U}_j)} - 1}} \quad (E1k)$$

Finally, we provide some approximations.
When the maximum uninhibited firing frequency $\bar{U}^*$ is positive, that alternative will dominate and the sum of the MI firing firing frequencies will be close to the uninhibited firing frequency of the maximum. That is, $m = \bar{U}^* + \dfrac{\omega}{\lambda}$, with $\omega$ small. Then, by definition
$$e^{-\lambda \bar{U}^* - \omega} = \prod \left(1 - e^{-\lambda(\bar{U}^* - \bar{U}_j)} e^{-\omega}\right) \quad (E1l)$$
$$e^{-\lambda \bar{U}^*} = e^{\omega} \prod \left(1 - e^{-\lambda(\bar{U}^* - \bar{U}_j)} e^{-\omega}\right) = (e^{\omega} - 1) \prod_{j \neq i^*} \left(1 - e^{-\lambda(\bar{U}^* - \bar{U}_j)} e^{-\omega}\right) \quad (E1m)$$
If $\omega$ is small, $e^{\omega} - 1 \cong \omega$ and $e^{-\omega} \cong 1 - \omega$, and $\omega^2 \cong 0$. It is found that:

$$e^{-\lambda \bar{U}^*} \cong \omega \prod_{j \neq i^*}\left(1 - e^{-\lambda(\bar{U}^* - \bar{U}_j)}\right) \quad (E1n)$$

Therefore, in that case, probabilities can be approximated by:

$$\begin{cases} \omega \cong \dfrac{e^{-\lambda \bar{U}^*}}{\prod_{j \neq i^*}\left(1 - e^{-\lambda(\bar{U}^* - \bar{U}_j)}\right)} > 0 \\ P_i \cong \dfrac{\ln\left(1 - e^{-\lambda(\bar{U}^* - \bar{U}_i)} e^{-\omega}\right)}{\sum \ln\left(1 - e^{-\lambda(\bar{U}^* - \bar{U}_j)} e^{-\omega}\right)} \end{cases} \quad (E1o)$$

When variability approaches zero the probability distribution centers at the maximum. At the extreme where variability approaches infinity all alternatives have the same probability.

If, on the contrary, the maximum uninhibited firing frequency $\bar{U}^*$ is negative, then we expect MI firing frequencies to be close to zero. Thus, the sum of all, $m$, will also be close to zero. Then, in this case we can approximate probabilities as:

$$P_i \cong \frac{\ln\left(1 - e^{\lambda \bar{U}_i}\right)}{\sum \ln\left(1 - e^{\lambda \bar{U}_j}\right)} \quad (E1p)$$

In the previous formula the absolute value was added for clarity. It is noticed that when uninhibited firing frequencies are negative enough, so that the Taylor expansion of $\ln\left(1 - e^{\lambda \bar{U}_i}\right) \approx e^{\lambda \bar{U}_i}$ around zero is reasonable, then McFadden's formula is recovered.

## *Appendix F: Hick's law*

**Theorem 10:** First, just as in Theorem 9, define $\bar{U}_i$ as the expected value of uninhibited firing frequency, $\bar{f}_i$ as the expectation of the resulting MI firing frequency of $f_i$ and $\varepsilon_i \equiv (U_i - \bar{U}_i) - \sum_{j \neq i}(f_j - \bar{f}_j)$, which distributes logistic with mean zero and constant scale parameter $\lambda$. Then, as in Theorem 9, each neuron in the MI mechanism complies with:

$$\bar{f}_i = \frac{1}{\lambda} \ln\left(1 + e^{\lambda(\bar{U}_i - \sum_{j \neq i} \bar{f}_j)}\right) \quad (F1a)$$

Now assume there are N alternatives and all of them are ex-ante similar, so that their expected uninhibited firing frequency is $\bar{U}$ for all. Then their expected MI firing frequency $\bar{f}$ is also the same and:

$$\bar{f} = \frac{1}{\lambda} \ln\left(1 + e^{\lambda(\bar{U} - (N-1)\bar{f})}\right) \quad (F1b)$$

Re writing that equation we find:

$$e^{\lambda \bar{f}} = 1 + e^{\lambda \bar{U}} e^{-\lambda(N-1)\bar{f}} \quad (F1c)$$

Finally, implicitly differentiating with respect to the number of alternatives, we find:

$$\lambda e^{\lambda \bar{f}} \frac{d\bar{f}}{dN} = e^{\lambda \bar{U}} \lambda e^{-\lambda(N-1)\bar{f}} \left((N-1)\frac{d\bar{f}}{dN} + \bar{f}\right) \quad (F1d)$$

So that:

$$\frac{d\bar{f}}{dN} = -\frac{e^{\lambda(\bar{U}-N\bar{f})}}{1+e^{\lambda(\bar{U}-N\bar{f})}(N-1)}\bar{f} = -\frac{\left(1-e^{-\lambda\bar{f}}\right)}{e^{-\lambda\bar{f}}+\left(1-e^{-\lambda\bar{f}}\right)N}\bar{f} \quad (F1e)$$

Where the last equality follows from $e^{\lambda(\bar{U}-N\bar{f})} = 1 - e^{-\lambda\bar{f}}$

Now, since all alternatives have the same expected MI firing frequency, and the total time is the time needed to create a description of the problem at hand and the inverse of the frequency needed to reach that decision (the period, in units of time):

$$T = \varepsilon_0 + \frac{1}{\bar{f}} \quad (F1f)$$

Differentiating with respect to the number of alternatives, we find

$$\frac{dT}{dN} = -\frac{1}{\bar{f}^2}\frac{d\bar{f}}{dN} \quad (F1g)$$

And therefore;

$$\frac{dT}{dN} = \frac{1}{\bar{f}}\frac{\left(1-e^{-\lambda\bar{f}}\right)}{e^{-\lambda\bar{f}}+\left(1-e^{-\lambda\bar{f}}\right)N} \quad (F1h)$$

Or equivalently,

$$\frac{dT}{dN} = \frac{1}{\bar{f}}\frac{\partial}{\partial N}\ln\left(e^{-\lambda\bar{f}}+\left(1-e^{-\lambda\bar{f}}\right)N\right) \quad (F1i)$$

That is, the total change in time-response when varying the number of alternatives equals the partial derivative of a logarithmic function of the number of alternatives.

Pan, X. et al. (2014). "Reward inference by primate prefrontal and striatal neurons". J. Neurosci. 34, 1380–1396.

Plassmann, H., O'Doherty, J. & Rangel, A. (2007) "Orbitofrontal Cortex Encodes Willingness to Pay in Everyday Economic Transactions" The Journal of Neuroscience, vol. 27, nº 37, p. 9984 –9988.

Proctor RW, Schneider DW (2018). "Hick's law for choice reaction time: A review." Q J Exp Psychol (Hove). Jun;71(6):1281-1299. doi: 10.1080/17470218.2017.1322622.

Quartz, S., & Sejnowski, T. (1997) "The neural basis of cognitive development: A constructivist manifesto", Behavioral and Brain Sciences, 20(4), 537-556.

Rand, D. G. et al. (2014). "Social heuristics shape intuitive cooperation". Nat. Commun. 5, 3677.

Rand, D. G., Greene, J. D. & Nowak, M. A. (2012). "Spontaneous giving and calculated greed". Nature 498, 427–430.

Rangel, A., Camerer C., & Montague, P. (2008) "A framework for studying the neurobiology of value-based decision making" Nature reviews, Neuroscience, vol. 9, pp. 1-13.

Ratcliff, R. (1978). "A theory of memory retrieval". Psychol. Rev. 85, 59–108.

Ratcliff, R. & McKoon, G. (2008) "The Diffusion Decision Model: Theory and Data for Two-Choice Decision Tasks", Neural Computation, 20, 4, April, pp. 873 – 922.

Robbins, Lionel (1935) *An Essay on the Nature & Significance of Economic Science*, second edition. London: Macmillan.

Roitman, J.D. and Shadlen, M.N. (2002). "Response of neurons in the lateral intraparietal area during a combined visual discrimination reaction time task". J. Neurosci. 22, pp. 9475–9489.

Rosch, S. G., Lane, J. A. S., Samuelson, C. D., Allison, S. T. & Dent, J. L (2000). "Cognitive load and the equality heuristic: a two-stage model of resource overconsumption in small groups". Organ. Behav. Hum. Decis. Processes 83, 185–212.

Rustichini, Aldo (2005) "Neuroeconomics: Present and future", Games and Economic Behavior, Elsevier, vol. 52(2), August, pages 201-212.

Saavedra, J. (2013) "It's all in the Brain" CESifo Newsletter, No. 4424.

Samejima, K., Ueda, Y., Doya K. & Kimura, M. (2005) "Representation of action-specific reward values in the striatum" Science, nº 310, p. 1337–1340.

Savage, Leonard (1954) *The Foundations of Statistics*, John Wiley & Sons, Inc.

Schall, J. (2003) "Neural Correlates of Decision Processses: Neural and Mental Chronometry" Current Opinion in Neurobiology, vol. 13, nº 2, pp. 182-186.

Schwitzgebel, Eric (2008) "The Unreliability of Naïve Introspection", The Philosophical Review, 117 (2): 245–273.

Shrestha, A and Mahmood, A (2019). "Review of Deep Learning Algorithms and Architectures," in IEEE Access, vol. 7, pp. 53040-53065.

Shikri, O., Hansel, D. & Sompolinsky, H. (2003) "Rate models for conductance-based cortical neuronal networks", Neural Computation 5: 1809–1841.

Stein RB (1965). "A theoretical analysis of neuronal variability". Biophysical Journal 5:173–194.